\documentclass[12pt]{spieman}  
\usepackage{amsmath,amsfonts,amssymb}
\usepackage{graphicx}
\usepackage{setspace}
\usepackage{tocloft}
\usepackage{lineno}
\usepackage{subcaption}

\title{Differential Double Blind Fourier Holography (diff-DBFH)}

\author[a,*]{Oren Pedatzur}
\author[a]{David T. Chuss}
\affil[a]{Villanova University, Department of Physics, 800 E. Lancaster Ave, Villanova, 19085, PA, USA}

\cftpagenumbersoff{figure}
\cftpagenumbersoff{table} 
\begin{document} 
\maketitle

\begin{abstract}
High-performance optical systems, including space telescopes, require wavefront-sensing and correction capabilities to achieve resolution near the diffraction limit. Existing wavefront-sensing approaches span a broad range of architectures, but many require dedicated interferometric hardware, rely on computationally intensive nonlinear or iterative reconstruction algorithms or provide only limited wavefront information. Focal-plane wavefront-sensing methods are particularly attractive because they can operate without dedicated metrology hardware, and among these, Double Blind Fourier Holography (DBFH) stands out as a robust and computationally light alternative to nonlinear and iterative solvers. Here we introduce differential Double Blind Fourier Holography (diff-DBFH), a two-image variant of DBFH based on a small pupil-plane amplitude perturbation. Inspired by differential Optical Transfer Function (dOTF), diff-DBFH leverages the linear framework of DBFH to produce a more precise wavefront estimate. We derive the diff-DBFH formalism, compare its performance against dOTF using numerical simulations on a segmented aperture geometry modeled after the Habitable Worlds Observatory, analyze its sensitivity under noisy conditions, and demonstrate the method experimentally using an optical bench model of the same aperture.
\end{abstract}

\keywords{optics, wavefront sensing, segmented telescopes, space telescopes, phase retrieval, Habitable Worlds Observatory}

{\noindent \footnotesize\textbf{*} Oren Pedatzur, \linkable{orenpedatzur@gmail.com} }

\begin{spacing}{1}   

\section{Introduction}
\label{sect:intro}

We characterize an optical system as `diffraction limited' if it achieves resolution which is only limited by the size of the aperture collecting the light and its wavelength. Optimal imaging systems are designed to achieve performance as close as possible to the diffraction limit by minimizing wavefront aberrations and shaping a spherically converging (or diverging) wavefront. Optical design tries to minimize the fixed wavefront errors, but in many important systems the medium around the optical system may not have fixed optical properties, or the optical system components themselves may not be stationary. Common examples are earth-based telescopes or free-space laser communication systems which suffer from atmospheric turbulence, and segmented primary mirror space telescopes which even though are not subject to atmospheric aberrations, are designed to have numerous controllable degrees of freedom. In all of these examples, the ability to sense and control the wavefront (WFS\&C) is key to achieving near diffraction-limited performance.

The importance of this problem is evident from the myriad of solutions that are employed to tackle it in its various embodiments. Where deploying auxiliary optical hardware is acceptable, common solutions include Shack-Hartmann sensors \cite{chanan1986segment}, grisms \cite{feinberg2007trl}, non-redundant sparse aperture interferometry \cite{cheetham2012fizeau}, holographic dispersed fringe sensors  \cite{haffert2022phasing}, etc. A major challenge is presented when, by design, minimal or no additional optical hardware is allowed to be used. For example, considering a space telescope setting, an ideal WFS solution probes the aperture wavefront using only a few focal plane images obtained with the existing science instrument sensor. The `gold standard' method for segmented space telescopes, called Phase Diversity (PD), captures multiple focal plane images augmented by several known phase masks (often 5 weak de-focusing values) \cite{gonsalves1982phase,gonsalves2018phase,dean2003diversity,dean2006phase}. A non-linear inverse solver is then used to estimate the pupil wavefront common to all images. Phase diversity is extremely successful in part because it embodies an optimal compromise between additional hardware, computational complexity and precision. On the other end of the space of solutions we find the elegant differential Optical Transfer Function (dOTF) pioneered by Codona et al. \cite{codona2012theory,codona2013differential}. In this technique an approximation of the wavefront is obtained via the Fourier transform of the difference between two focal plane images: one of the full aperture, and one in which a small mask (blocker) is applied to the edge of the pupil that serves as an amplitude perturbation.
Apart from using a mere two images, dOTF's appeal is in its simple application and even simpler readout. The tradeoff is reduced phase estimation precision due to the perturbative approximation. Recent dOTF studies have used bisymmetric pupil modifications to reduce reconstruction blur \cite{martinez2023bisymmetric}, explored broadband wavefront sensing with integral-field spectroscopy \cite{martinez2024differential}, and demonstrated wavefront sensing from coronagraphic images \cite{martinez2025revisiting}.

Double Blind Fourier Holography is a phase retrieval method first introduced by Raz et al. for 1D signals \cite{raz2011vectorial,raz2013vectorial} and later applied to lensless imaging \cite{raz2014direct}. It has since been used to solve several key phase retrieval challenges, including X-ray free-electron laser (XFEL) coherent diffraction imaging of nano-particles \cite{leshem2016direct} and attosecond pulse characterization \cite{pedatzur2019double}. Recently, it has been proposed as a candidate solution to the extreme picometer precision WFS requirement of the Habitable Worlds Observatory (HWO) \cite{pedatzur2026double}, a future space telescope whose goals include directly imaging faint exoplanets in the vicinity of bright stars. Experimental validation of DBFH for wavefront sensing has demonstrated Technology Readiness Level 3 (TRL 3) \cite{pedatzur2026experimental}. In the context of WFS, DBFH requires a set of three focal plane measurements corresponding to full aperture and two mutually exclusive and complementary sub-apertures. A linear combination of these focal plane measurements provides an estimate to the interference cross term. The estimated cross-term, together with the apertures' region of compact support (CS, the mathematically closed and bounded set of points where the illumination is non-zero), constitute an overdetermined system of linear equations whose solution is the complex field in the focal plane, which is then back-propagated to recover the pupil plane wavefront. DBFH presents an improvement over existing methods such as phase diversity in terms of significantly lower computational load, uniqueness of solution owing to its truly algebraic linear form, and its lower number of required captures and dedicated optical hardware. 
Rather than introducing defocus, DBFH encodes the wavefront information through known amplitude masking in the pupil plane. This makes it particularly attractive for precision imaging and astrometry use-cases, where deliberate phase modulation and optical-path perturbations are undesirable.

In this work we present, validate and explore differential DBFH (diff-DBFH), a novel WFS method inspired by both DBFH and dOTF, and which fuses key elements from both. Similar to dOTF, diff-DBFH subtracts two focal plane images of the full pupil and a perturbed aperture and neglects the intensity associated with the blocked field to approximate the interference cross-term. The approximate cross-term and the known CS are then mapped to the required DBFH algorithm inputs. Diff-DBFH trades the simple autocorrelation-space readout used in dOTF with the linear-system solution at the heart of DBFH, substantially improving wavefront-estimation precision.

The goal of this paper is to introduce and establish diff-DBFH as a fast, perturbative, and computationally light WFS method which provides a better wavefront estimate relative to dOTF. Section \ref{section: Differential DBFH} reviews the mathematical description, assumptions, requirements, algorithmic steps and limitations associated with diff-DBFH. Section \ref{section: numerical study} presents a numerical study comparing diff-DBFH and dOTF over various noise regimes and different perturbation geometries, on the Exploratory Analytical Case \#5 (EAC5) telescope model proposed for the Habitable Worlds Observatory (HWO) \cite{feinberg2026habitable}. In section \ref{section: experimental results} we provide a simple experimental demonstration of diff-DBFH, and compare its recovery to that of dOTF with identical captures, and a reference solution provided by the full DBFH. We present our conclusions to this study in section \ref{section: conclusions}. Finally, we provide accessible open-source code that enables users ranging from observatory optical engineers to amateur astronomers to readily implement and adapt diff-DBFH for diverse wavefront-sensing applications.

\section{Differential DBFH} \label{section: Differential DBFH}

\subsection{Mathematical Description of the differential DBFH Algorithm} \label{subsection: diff-DBFH math}
The complex electric field at the pupil plane $E(\mathbf{x})$ of an infinity conjugate imaging system (e.g., a telescope) is related to the field at the focal plane $ A(\boldsymbol{\xi})$ by a 2D Fourier transform:
\begin{equation}
    A(\boldsymbol{\xi}) = \mathcal{F}[E(\mathbf{x})],
\end{equation}
where $\mathbf{x}$ and $\boldsymbol{\xi}$ are pupil plane and focal plane coordinates correspondingly.
A sensor placed in the focal plane will record an intensity image
\begin{equation}
    I(\boldsymbol{\xi}) = |A(\boldsymbol{\xi})|^2.
\end{equation}
Focal plane wavefront sensing methods, such as phase-diversity, DBFH, diff-DBFH and dOTF, seek to characterize $\phi(\mathbf{x})=\arg(E(\mathbf{x}))$ by studying a non-degenerate set of focal plane images.

We consider a known small localized blocking mask $b(\mathbf{x})$ (as done in dOTF), such that
\begin{alignat*}{2}
    & b(\mathbf{x})=0 \quad \text{inside a small sub-aperture } B, \\
    & b(\mathbf{x})=1 \quad \text{elsewhere.}
\end{alignat*}

We decompose the full pupil field into two disjoint spatial components:
\begin{equation}
    E(\mathbf{x}) = E_1(\mathbf{x}) + E_2(\mathbf{x}),
\end{equation}
where $E_2(\mathbf{x})$ is supported on $B$, and $E_1(\mathbf{x})$ is supported on the complementary exposed pupil. By construction, $E_1$ and $E_2$ have non-overlapping compact supports.
In the focal plane we define
\begin{gather}
    A_1(\boldsymbol{\xi}) = \mathcal{F}[E_1(\mathbf{x})],\\
    A_2(\boldsymbol{\xi}) = \mathcal{F}[E_2(\mathbf{x})],\\
    A(\boldsymbol{\xi}) = \mathcal{F}[E(\mathbf{x})] = A_1(\boldsymbol{\xi}) + A_2(\boldsymbol{\xi}).
\end{gather}
Engaging the blocker removes the contribution of $E_2$, so the blocked-pupil image is $I_1=|A_1|^2$, whereas the full-pupil image is $|A_1 + A_2|^2$. Their difference is therefore
\begin{equation}
    \Delta I(\boldsymbol{\xi})
    = |A_1(\boldsymbol{\xi}) + A_2(\boldsymbol{\xi})|^2 - |A_1(\boldsymbol{\xi})|^2 
    = |A(\boldsymbol{\xi})|^2 - |A_1(\boldsymbol{\xi})|^2
    = 2\operatorname{Re}\!\left(A_1^*(\boldsymbol{\xi}) A_2(\boldsymbol{\xi})\right)
    + | A_2(\boldsymbol{\xi})|^2.
\end{equation}
In full (non-perturbative) DBFH, $|A_1|$ and $|A_2|$ are comparable, and the quadratic term $I_2(\boldsymbol{\xi}) = |A_2(\boldsymbol{\xi})|^2$ is measured in a third image and subtracted from $\Delta I$. The small blocked area motivates a first-order approximation in the perturbation field for diff-DBFH, in which we neglect $I_2(\boldsymbol{\xi})$, yielding
\begin{equation}
    \Delta I(\boldsymbol{\xi})
    \approx
    2\operatorname{Re}\!\left(A_1^*(\boldsymbol{\xi}) A_2(\boldsymbol{\xi})\right).
\end{equation}
This approximation is justified where the quadratic term is much smaller than the magnitude of the cross term, but can become inaccurate near zeros of the cross term, even for a weak perturbation.

The difference between the focal plane intensity captures provides an approximation for $\operatorname{Re}\!\left(A_1^* A_2\right)$, but in order to apply DBFH we need the complex field corresponding to $A_1^* A_2$. Subsection \ref{subsec: Im(A_1^*A_2)} details a procedure by which the imaginary part of a complex function can be estimated, provided we know its real part and that its Fourier transform vanishes over a half-plane.

\begin{figure}[H]
    \centering
    \includegraphics[width=0.9\linewidth]{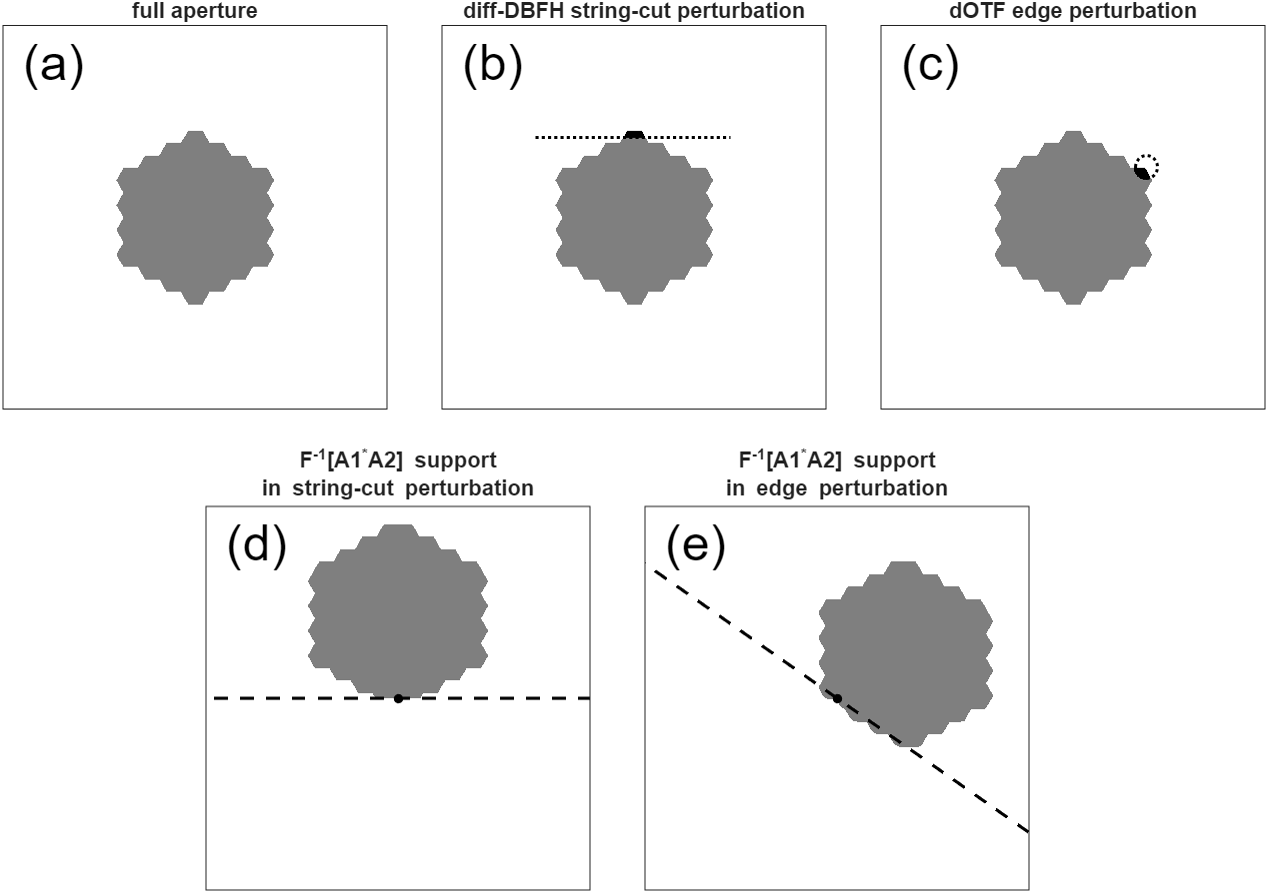}
    \vspace{1mm}
    \caption{\textbf{Pupil and perturbation geometries.}  Comparison of two pupil-plane perturbation geometries for the HWO EAC5-like aperture. (a) Full aperture. (b) A straight-line `string-cut' perturbation used for diff-DBFH, dotted line. (c) An edge-localized perturbation typically used for dOTF, defined by a section of the dotted circle whose center is on the edge of the aperture. The corresponding supports of $g=\mathcal{F}^{-1}[A_1^*A_2]$ are shown for the `string-cut' (d) and `edge' (e) perturbations. Recovery of $\operatorname{Im}(A_1^*A_2)$ requires the support of $g$ to lie entirely within one half-plane bounded by a line passing through the origin. The straight-cut geometry satisfies this condition, as shown in (d). For the edge perturbation considered in (c) the support extends to both sides of any candidate line through the origin, and therefore does not satisfy the required half-plane support condition. The black dot denotes the origin, and the dashed lines represent candidate half-plane boundaries.}
    \label{fig:perturbation_cuts}
\end{figure}

We note that not every choice of perturbation geometry, $B$, satisfies the above requirement that  $\mathcal{F}^{-1}[A_1^* A_2]$ vanishes over a half-plane. Specifically, for the perturbations used by Codona \cite{codona2013differential}, disk or finger-like obscuration at the edge of the pupil, the support of the convolution $E_1^*(\mathbf{-x})* E_2(\mathbf{x})$ extends into both sides of every choice of half-plane passing through the origin. By contrast, a simple straight line cut between $E_1,E_2$ satisfies this condition trivially. Figure \ref{fig:perturbation_cuts} compares the pupil-plane support of $\mathcal{F}^{-1}[A_1^* A_2]$ for both Codona's perturbation and a straight line cut. 

Following the procedure detailed in subsection \ref{subsec: Im(A_1^*A_2)}, we obtain a complex estimate for the cross-term:
\begin{equation}
    C(\boldsymbol{\xi}) = A_1^*(\boldsymbol{\xi}) A_2(\boldsymbol{\xi}).
\end{equation}
and we are ready to construct the required equations for DBFH. Adopting a phasor notation for the unknown focal plane phases $X_1(\boldsymbol{\xi})=e^{i\phi_1(\boldsymbol{\xi})}$ and $X_2(\boldsymbol{\xi})=e^{i\phi_2(\boldsymbol{\xi})}$, we decompose the focal plane fields into the corresponding amplitudes and phasors: $A_1 = |A_1|X_1$ and $A_2 = |A_2|X_2$. We multiply $C(\boldsymbol{\xi})$ by the unknown phasor $X_1$, to get a first linear relationship between the phases of $A_1$ and $A_2$:
\begin{equation}
    C(\boldsymbol{\xi})X_1=|A_1(\boldsymbol{\xi}) A_2(\boldsymbol{\xi})|X_2
      \label{eq: crossterm}
\end{equation}
We also require that both fields $E_1,E_2$ vanish outside of their respective regions of CS (at the pupil plane):
\begin{gather}
    \mathcal{F}^{-1}\!\left[\sqrt{I_1(\boldsymbol{\xi})} X_1(\boldsymbol{\xi})\right]
    = 0 \qquad \text{for } \mathbf{x} \notin CS_1\\
    \mathcal{F}^{-1}\!\left[\sqrt{I_2(\boldsymbol{\xi})} X_2(\boldsymbol{\xi})\right]
    = 0 \qquad \text{for } \mathbf{x} \notin CS_2  \label{eq: CS2}
\end{gather}
As we do not have a direct measurement of $I_2(\boldsymbol{\xi})$ for eq. \ref{eq: CS2}, we might be tempted to substitute it using $C(\boldsymbol{\xi})$ from eq. \ref{eq: crossterm}:
\begin{equation}
    \sqrt{I_2(\boldsymbol{\xi})}X_2 = \frac{C(\boldsymbol{\xi})}{\sqrt{I_1(\boldsymbol{\xi})}}X_1;
\end{equation}
however, this risks amplifying noise in pixels with low values of $\sqrt{I_1}$. We've found that solving for the unknown field, $A_2$, instead of its phasors, $X_2$, formulates a more tractable linear problem.

Finally, we obtain the DBFH system of linear equations for the unknown fields $X_1$, and $A_2$:
\begin{gather}
    \mathcal{F}^{-1}\!\left[\sqrt{I_1(\boldsymbol{\xi})} X_1(\boldsymbol{\xi})\right]
    = 0 \qquad \text{for } \mathbf{x} \notin CS_1,\\
    \mathcal{F}^{-1}\!\left[A_2(\boldsymbol{\xi})\right]
    = 0 \qquad \text{for } \mathbf{x} \notin CS_2,\\
    C(\boldsymbol{\xi})X_1(\boldsymbol{\xi})=|A_1(\boldsymbol{\xi})| A_2(\boldsymbol{\xi}).
\end{gather}

This homogeneous system is invariant under multiplication of both unknown fields by an arbitrary nonzero complex scalar. We remove this common scale and global-phase ambiguity by fixing one component of $X_1(\boldsymbol{\xi}) = 1$. Specifically, we select a well-measured focal-plane pixel to serve as an anchor, and impose the gauge condition $X_1(\boldsymbol{\xi}_{anchor}) = 1$. This converts the homogeneous system into a nonhomogeneous linear system, which we solve in the least-squares sense, recovering estimates for both $X_1$ and $A_2$. Because the unit-modulus constraint on $X_1$ is relaxed during the linear solve, the recovered phasor is finally projected onto the unit circle: 
\begin{equation}
    \hat X_1(\boldsymbol{\xi}) =  \frac{X_1(\boldsymbol{\xi})}{|X_1(\boldsymbol{\xi})|}
\end{equation}

A non-homogeneous, overdetermined linear system has a unique least-squares solution if its coefficient matrix has full column rank \cite{lay2015linear}. The constraints in our system arise from sampling points outside $CS_1$, sampling points outside $CS_2$, and pixels in the overlap of the point spread functions (PSF) $I_1(\boldsymbol{\xi})$ and $I_2(\boldsymbol{\xi})$ at the focal plane. The overlap is defined as pixels with intensity above the noise floor in both images. As the full and perturbative PSFs are very similar, this overlap is naturally large. Oversampling the focal plane above the Nyquist limit increases the number of available constraints, although their number alone does not establish full column rank. In the presence of noise, the system is generally inconsistent and is therefore solved in the least-squares sense. The recovered phase is then unique and defined modulo $2\pi$, as in any monochromatic phase-retrieval method.

\subsection{Recovering $\operatorname{Im}(A_1^*A_2)$ using the Slice--Projection theorem} 
\label{subsec: Im(A_1^*A_2)}
We consider the problem of recovering the imaginary part of the cross term,
\begin{equation}
    z(\boldsymbol{\xi}) = A_1^*(\boldsymbol{\xi}) A_2(\boldsymbol{\xi}),
\end{equation}
given a measurement of its real part $\operatorname{Re}[z(\boldsymbol{\xi})]$ and prior knowledge that its inverse Fourier transform is compactly supported and vanishes over a half-plane. 
As we saw in subsection \ref{subsection: diff-DBFH math}, this problem arises in diff-DBFH when decomposing the pupil field into two disjoint regions: $E_1(\mathbf{x})$ the main aperture (excluding the perturbative region), and $E_2(\mathbf{x})$, the small perturbative region (covered by a blocker). The corresponding focal-plane fields are $A_1$ and $A_2$, and the cross term is $z = A_1^* A_2$. This problem in the context of DBFH was introduced and solved by Raz et al. for rectangular apertures \cite{raz2014direct}, but the solution is not directly transferable to arbitrary aperture shapes.

The 2D inverse Fourier transform of the complex cross-term can be expressed as a convolution of the two disjoint pupil plane fields:
\begin{equation}
    g(\mathbf{x}) = \mathcal{F}^{-1}\{z(\boldsymbol{\xi})\}=\mathcal{F}^{-1}[A_1^*] * \mathcal{F}^{-1}[A_2]
\end{equation}
This term is defined in the autocorrelation domain, and since
\begin{equation}
    \mathcal{F}^{-1}[A_1^*](\mathbf{x}) = E_1^*(-\mathbf{x}), 
    \qquad 
    \mathcal{F}^{-1}[A_2](\mathbf{x}) = E_2(\mathbf{x}),
\end{equation}
we obtain
\begin{equation}
    g(\mathbf{x}) = E_1^*(-\mathbf{x}) * E_2(\mathbf{x}).
\end{equation}
Therefore, the support of $g$ is contained within the set of all displacement vectors,
\begin{equation}
    \mathbf{x} = \mathbf{x}_2 - \mathbf{x}_1,
    \quad
    \mathbf{x}_1 \in CS_{1},\;
    \mathbf{x}_2 \in CS_{2}.
\end{equation}
A \emph{straight-cut geometry} is defined using a normal to the cut $\hat{\mathbf{n}}$ , and a constant $c$ such that: 
\begin{equation}
    \mathbf{x_1} \cdot \hat{\mathbf{n}} \leq c ~ \text{for all} ~\mathbf{x_1} \in CS_{1}
\end{equation}
and 
\begin{equation}
    \mathbf{x_2} \cdot \hat{\mathbf{n}} \geq c ~ \text{for all} ~\mathbf{x_2} \in CS_{2}
\end{equation}
Hence, $\hat{\mathbf{n}}\cdot(\mathbf{x_2}-\mathbf{x_1}) \geq 0$, or equivalently, for any choice of $(\mathbf{x_1},\mathbf{x_2})$, the point $\mathbf{x_2}-\mathbf{x_1}$ must lie on the positive side defined by the normal $\hat{\mathbf{n}}$. The support of $g(\mathbf{x})$ is included in this half-plane
\begin{equation}
    CS_g \subseteq \{\mathbf{x}:\hat{\mathbf{n}}\cdot\mathbf{x}\geq 0\}
\end{equation}
This means that the function $g(\mathbf{x})$ must vanish in the half-plane defined by $\hat{\mathbf{n}}\cdot\mathbf{x} < 0$. 
An example to the support of $g(\mathbf{x}) =\mathcal{F}^{-1}[A_1^*A_2]$ in geometry corresponding to the future HWO space telescope can be seen in fig \ref{fig:perturbation_cuts}.

Next, we employ the projection-slice theorem to analyze $g(\mathbf{x})$ in polar coordinates. Let $\theta$ define a direction and consider $g_\theta(s) = g(s\,\boldsymbol{\hat{\theta}})$, a 1D slice sampling the function $g(\mathbf{x})$ and passing through the origin (center) \footnote{In our choice of polar coordinates, $\theta\in [ 0,\pi )$ and $s \in(-\infty,\infty)$. This is convenient for specifying half-plane regions using the sign of $s$.}. By the projection--slice theorem,
\begin{equation}
    g_\theta(s) = \mathcal{F}^{-1}_{1D}\{ R_\theta[z] \}
\end{equation}
where $R_\theta[z]$ is the Radon transform (projection) of $z$ along direction $\boldsymbol{\hat{\theta}}$. To avoid confusion, we distinguish 1D Fourier transforms using the notation $\mathcal{F}_{1D}[\,~\,]$. Decomposing $z$ to its real and imaginary parts, $z = \operatorname{Re}\,z + i\,\operatorname{Im}\,z$, we obtain
\begin{equation}
    R_\theta[z] = R_\theta[\operatorname{Re}\,z] + i\,R_\theta[\operatorname{Im}\,z]
\end{equation}
Thus,
\begin{equation}
    g_\theta(s) = g_{R,\theta}(s) + i\,g_{I,\theta}(s)
    \label{eq: g_theta}
\end{equation}
with
\begin{align}
    g_{R,\theta}(s) &= \mathcal{F}^{-1}_{1D}\{R_\theta[\operatorname{Re}\,z]\} \\
    g_{I,\theta}(s) &= \mathcal{F}^{-1}_{1D}\{R_\theta[\operatorname{Im}\,z]\}
\end{align}
Since $\operatorname{Re}\,z$ is measured, $g_{R,\theta}(s)$ is known.
From the half-plane support condition,
\begin{equation}
    g_\theta(s) = 0 \quad \text{for } s < 0
\end{equation}
therefore, from eq. \ref{eq: g_theta}
\begin{equation}
    g_{I,\theta}(s) = i\,g_{R,\theta}(s), \quad s < 0.
\end{equation}
Since $R_\theta[\operatorname{Im}\,z]$ is real-valued, its inverse Fourier transform satisfies Hermitian symmetry:
\begin{equation}
    g_{I,\theta}(-s) = g^*_{I,\theta}(s)
\end{equation}
Thus, knowledge of $g_{I,\theta}(s)$ for $s < 0$ determines it for $s > 0$, yielding $g_{I,\theta}(s)$ for all $s$.

Once $g_{I,\theta}(s)$ is known, we compute
\begin{equation}
    R_\theta[\operatorname{Im}\,z] = \mathcal{F}_{1D}\{g_{I,\theta}(s)\}.
\end{equation}
Thus, for each $\theta$, we obtain a projection of $\operatorname{Im}\,z$.

A standard tomographic inversion (i.e. inverse Radon transform using back projections), may be used to recover an estimate for $\operatorname{Im}\,z(\boldsymbol{\xi})$, but is unadvisable for such a delicate numerical task. We've found that common back projection algorithms which smear values from each projection onto the normal direction introduce biases which are detrimental to DBFH reconstruction. Instead, we interpolate $g_{I,\theta}(s)$ from polar coordinates back to a cartesian grid, and apply a 2D Fourier transform to recover $\operatorname{Im}\,z$.

A numerical example to the recovery of $\operatorname{Im}(A_1^*A_2)$ for our perturbation and aperture geometry can be found in Appendix A, section \ref{appendix: Estimation of $Im(A_1^*A_2)$}.

\section{Numerical Performance Study} \label{section: numerical study}

We use the Habitable Worlds Observatory (HWO) Exploratory Analytical Case 5 (EAC5) aperture as a representative segmented-aperture geometry for the numerical investigation. This geometry combines several features seen as relevant to future large space telescopes, including the non-trivial pupil structure and the hexagonal segmentation. The broader DBFH framework has previously been shown capable of picometer-scale wavefront sensitivity \cite{pedatzur2026double}, motivating its consideration for high-precision optical systems. However, the perturbative approximation in diff-DBFH introduces bias into the wavefront estimate. We therefore do not expect diff-DBFH to achieve the full picometer-level wavefront-sensing capability required for HWO. Rather, we use the HWO-like aperture as a benchmark geometry for evaluating the behavior of a linear, two-image focal-plane estimator and for comparing its performance with dOTF. In this context, diff-DBFH may be useful in less demanding regimes, such as coarse co-phasing of segmented apertures, rapid detection of misalignment, for establishing an initial wavefront error prior for higher-precision nonlinear or iterative reconstruction methods, or when deliberate defocus or other optical-path perturbations are undesirable.

Our simulation includes a $2048\times2048$ grid sampling the pupil plane and focal plane fields. We assume sensor pixel pitch of $px=3.45 \mathrm{\mu m}$, which together with central wavelength of $\lambda=550\mathrm{nm}$ and focal distance $f=171\mathrm{m}$, set the pupil plane simulation box size at $L_{pupil}=\frac{\lambda f}{px}=27.3\mathrm{m}$. Each sampling bin in the pupil plane corresponds to $13.3\,\mathrm{mm}$. The inscribed primary mirror diameter of $D=8.2\,\mathrm{m}$ spans $616$ pixels, while the flat-to-flat diameter of each hexagonal segment ($\sim1.17\,\mathrm{m}$) spans $88$ pixels. The full 37-segment aperture therefore has a total geometric area of approximately $2.48\times10^5$ pixels. These conditions give rise to a sampling parameter $Q=\frac{\lambda~f}{D ~px}\sim 3.3$ which satisfies the necessary oversampling condition for DBFH and the recovery of $\operatorname{Im}(A_1^*A_2)$.

\begin{figure}[H]
    \includegraphics[width=\linewidth]{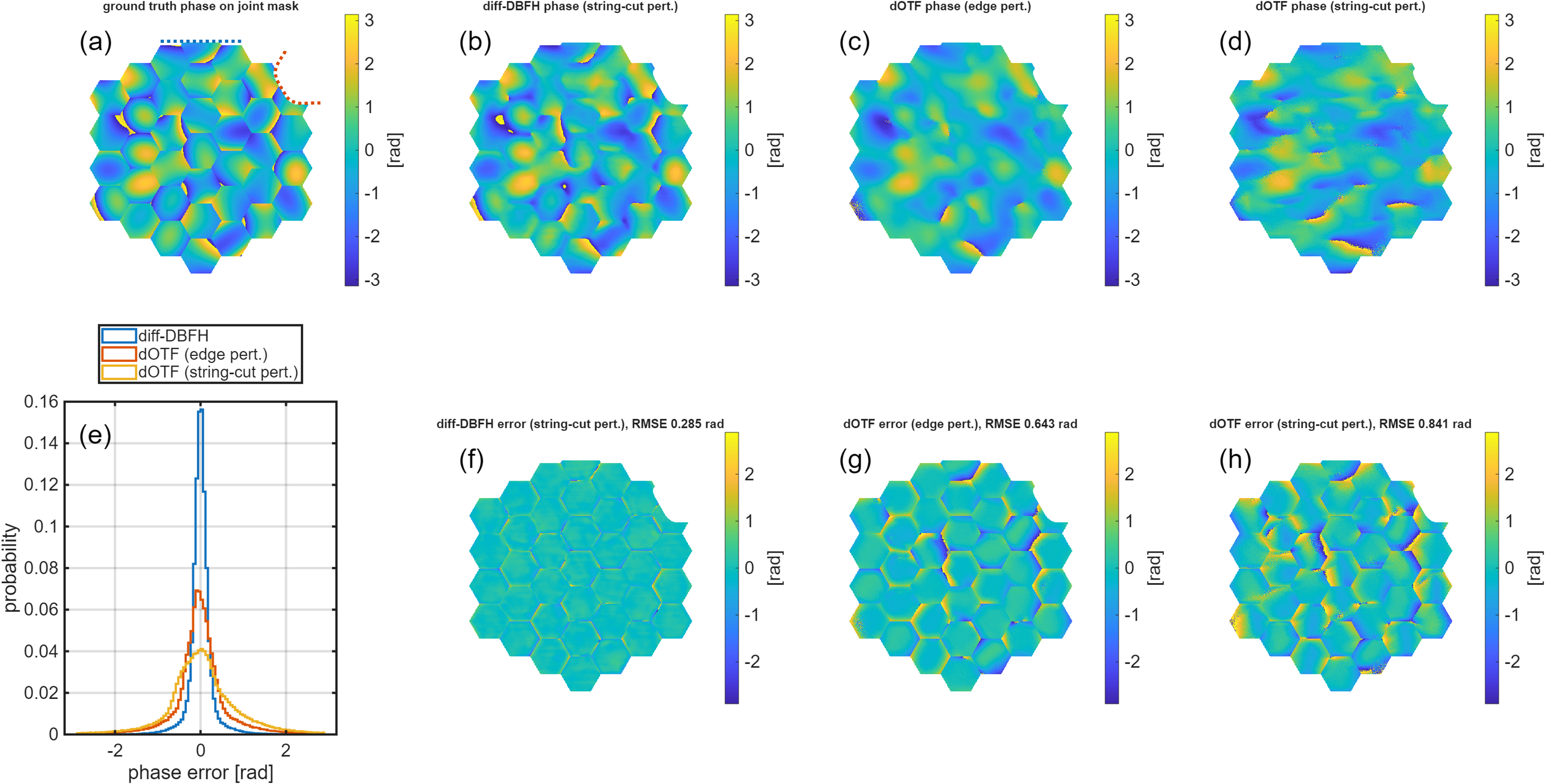}
    \vspace{1mm}
    \caption{\textbf{Numerical comparison.} Top row: (a) ground truth wavefront along with the `string-cut' (dashed blue) and `edge' (dashed orange) perturbation geometries. Wavefront phase estimation using diff-DBFH (b) and dOTF (c), and a complementary second implementation of dOTF using `string-cut' (d). Bottom row: (e) histogram and spatial distribution of errors using each method (f-h). The error analysis and all pupil plane figures exclude the union of perturbative regions where the respective methods are expected to perform poorly.}
    \label{fig:joint-mask phase error comparison}
\end{figure}

Over the above pupil geometry we generate a random wavefront error (WFE) which consists of random values for the 6 lowest Zernike coefficients ($n\leq2$ including piston, tip/tilt, defocus and astigmatism) for each hexagonal segment. Each non-piston mode is mean-subtracted and normalized to unit RMS over the segment. Their coefficients are drawn from a zero-mean and $\sigma=0.5\mathrm{rad}$ normal distribution, fig. \ref{fig:joint-mask phase error comparison}(a). We then set two perturbation geometries: a `string-cut' which is the only plausible choice for diff-DBFH, and a simple circular `edge' blocker which is expected to optimize dOTF. Importantly, to constitute a fair comparison, the areas of both perturbation choices are equal to exactly $959$ pixels. We propagate the fields to the focal plane using a Fourier transform, generate the intensity images that correspond to the full and perturbed apertures, and add an independent realization of Gaussian noise to each image, with standard deviation $\sigma_{noise}=10^{-5}$ relative to the peak intensity of the full-aperture image. Throughout this work, measured SNR denotes the total intensity within a region containing 95\% of the total image intensity, divided by the standard deviation of the summed noise in that region.

Next we apply diff-DBFH using the differential intensity corresponding to the `string-cut' perturbation, and dOTF to the differential intensity corresponding to the `edge' perturbation. In order to bridge the gap between these two different perturbation geometries we also apply dOTF using the `string-cut' input. Figures \ref{fig:joint-mask phase error comparison}(b-d) show the recovered wavefront phase in each modality. Any global phase between the recovered pupil fields, $E_{\mathrm{sol}}(\mathbf{x})$, and the ground-truth reference, $E_{\mathrm{ref}}(\mathbf{x})$, is estimated as a weighted circular mean over valid pupil pixels selected by the evaluation mask and then removed:
\begin{equation}
    \theta = \arg\!\left[\sum_{\mathbf{x}\in mask}
    E_{\mathrm{sol}}^*(\mathbf{x})E_{\mathrm{ref}}(\mathbf{x})\right],
    \quad
    E_{\mathrm{aligned}}(\mathbf{x}) = e^{i\theta}E_{\mathrm{sol}}(\mathbf{x}).
\end{equation}
We use a cyclic phase difference to quantify the reconstruction errors in each pupil plane coordinate: 
\begin{equation}
    \Delta\phi(\mathbf{x}) =\arg\!\left[E_{\mathrm{aligned}}(\mathbf{x})E_{\mathrm{ref}}^*(\mathbf{x})\right],
\end{equation}
The phase reconstruction error for a pupil mask containing $N$ samples is given by the root-mean-square of the phase difference:
\begin{equation}
    \phi_{\mathrm{RMSE}} = \sqrt{\frac{1}{N}\sum_{\mathbf{x}\in mask}\Delta\phi(\mathbf{x})^2}.
\end{equation}

Figures \ref{fig:joint-mask phase error comparison}(f-h) show the phase reconstruction error (relative to the ground-truth) corresponding to each of the perturbative methods. Figure \ref{fig:joint-mask phase error comparison}(e) summarizes these phase error results in a histogram showing a narrow distribution of $\phi_{RMSE}=0.285~\mathrm{rad}$ for diff-DBFH, and a significantly wider error distribution corresponding to dOTF with $\phi_{RMSE}=0.643~\mathrm{rad}$. The intermediate case of dOTF using the `string-cut' perturbation produced the worst results, as this choice of perturbation is sub-optimal as per dOTF. Again, to facilitate a fair comparison, the perturbation areas associated with low accuracy for each of the methods were disregarded for the purpose of this analysis (and are masked in the figures).

\begin{figure}[H]
    \includegraphics[width=\linewidth]{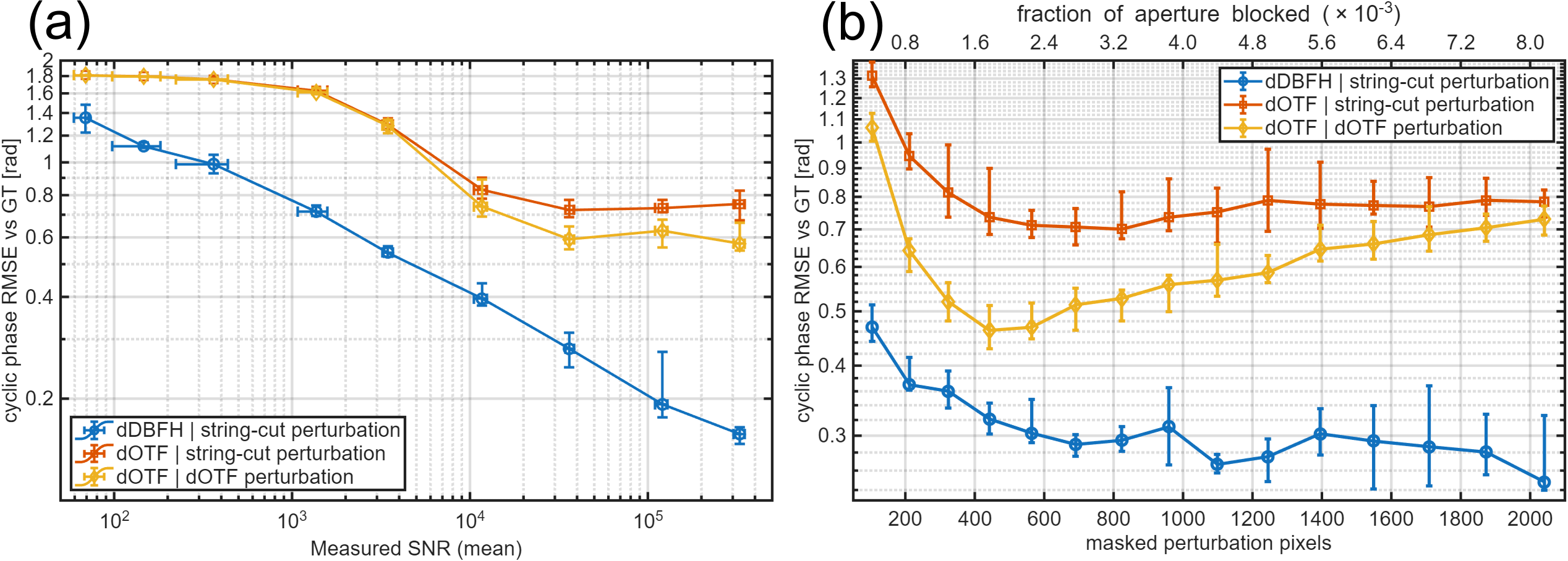}
    \caption{\textbf{Response to noise and perturbation size.} Each point summarizes 20 random phase and noise realizations at one simulation condition. In (a), phase RMSE relative to the simulated ground truth is plotted against the mean measured SNR at each noise level. Horizontal bars span the 25th to 75th percentiles of measured SNR. In (b), phase RMSE is plotted against perturbation area at fixed noise level. The bottom axis gives the blocked area in pixels, and the top axis gives the corresponding fraction of the full aperture area. Vertical bars in both panels span the 25th to 75th percentiles of phase RMSE across the 20 random phase and noise realizations.}
\label{fig:diff-DBFH vs dOTF performance}
\end{figure}

Both methods are expected to perform differently depending on the perturbation size and noise level. In short, a larger perturbation increases the signal strength (SNR, relative to noise) in the intensity difference image, but due to the worsening approximation, it also introduces a stronger bias. This noise and bias tradeoff in diff-DBFH is discussed further in Appendix B, \ref{appendix: Noise and Perturbation Size Tradeoff}. We compare diff-DBFH and dOTF in two numerical sweeps, first varying the noise level at fixed perturbation area and then varying the perturbation area at fixed noise level. At each point, we simulate 20 random phase realizations and add independent Gaussian noise to the full- and perturbed-aperture images to model separate exposures. Variations in the wavefront and noise produce different measured SNR values even at a fixed normalized noise level. Fig.~\ref{fig:diff-DBFH vs dOTF performance}(a) shows phase RMSE against measured SNR, averaged over the 20 realizations, for Gaussian noise levels  $\sigma_{\mathrm{noise}} \in \left[10^{-6},10^{-2}\right]$ relative to the full-aperture peak intensity at a fixed perturbation area of 1320 pixels. Fig.~\ref{fig:diff-DBFH vs dOTF performance}(b) shows phase RMSE against perturbation area at a fixed $\sigma_{\mathrm{noise}}=10^{-5}$, with a measured SNR of $(3.6\pm0.4)\times10^{4}$ across the sweep. Diff-DBFH yields lower phase RMSE than dOTF throughout both sweeps. The increase in dOTF error at larger perturbation areas may partly reflect a mechanism specific to the dOTF readout: convolution with the finite perturbation field blurs spatial detail in the recovered pupil field.

\section{Experimental Demonstration} \label{section: experimental results}

A basic experimental demonstration is useful both for completing the comparison between diff-DBFH and dOTF and for emphasizing the practical simplicity of pupil-plane amplitude masking, particularly in perturbative focal-plane methods such as diff-DBFH and dOTF. Full DBFH, previously established as a highly accurate wavefront-sensing method, is used here as the reference reconstruction against which both perturbative methods are evaluated.

\begin{figure}[H]
    \includegraphics[width=1\textwidth]{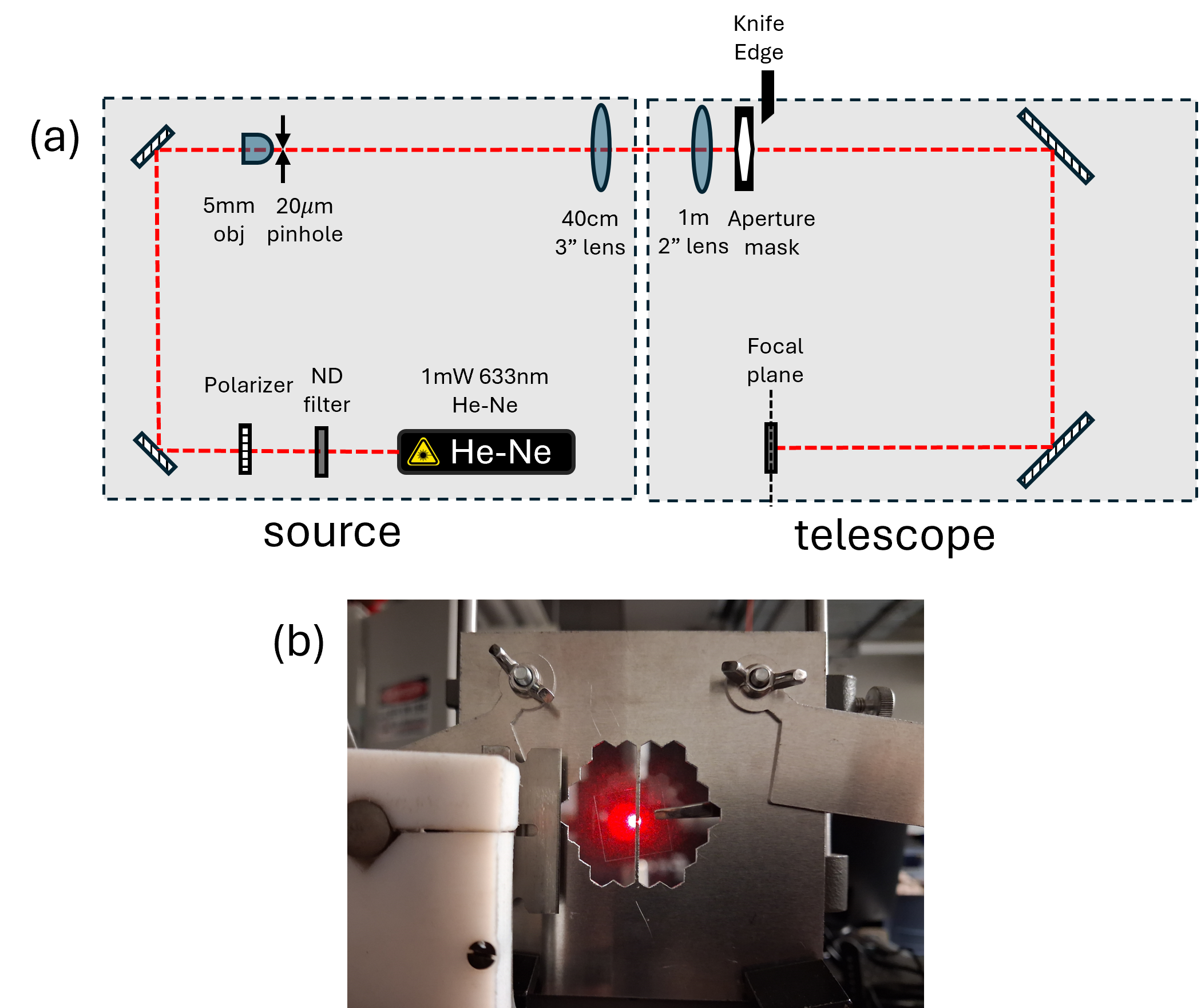}
    \label{fig:experimental_setup}
    \caption{\textbf{Experimental setup.} (a) Simple experimental setup used to capture DBFH, diff-DBFH and dOTF focal plane images. (b) A photograph of the pupil mask taken from the sensor side, showing also the horizontally adjustable knife edge (left), two manual shutters for DBFH, and a fixed microscope cover-slip used as a non-trivial phase element.}
\label{fig:exp_setup and photo}
\end{figure}

Our experimental bench setup, depicted in \ref{fig:exp_setup and photo}(a), is designed to model HWO EAC5 geometry with $Q\sim4$. The source is a $1\mathrm{mW}$ He-Ne laser, $\lambda = 633\mathrm{nm}$. The beam is first conditioned using a ND filter and a linear polarizer, and then spatially filtered and expanded using a $5\mathrm{mm}$ objective, $25\mathrm{\mu m}$ diameter pinhole and a $f=40\mathrm{cm},D=7.62\mathrm{cm}$ lens. A laser cut aperture mask of diameter $D=40\mathrm{mm}$, fig. \ref{fig:exp_setup and photo}(b), is placed right in front of the telescope primary focusing lens $f=1\mathrm{m},D=50\mathrm{mm}$. The aperture mask design includes two manual shutters to allow blocking or exposing the left and right halves of the aperture to provide inputs for the reference DBFH WFS. The `string-cut' perturbation is applied to the full aperture using a vertical knife-edge manually inserted into the beam using a micro-metric stage. A validation phase element, $200\mathrm{\mu m}$ glass microscope cover slip, is placed near the aperture using tweezers. The focal images are captured using a Zelux 1.6 MP Monochrome CMOS sensor (Thorlabs, $3.45 \mathrm{\mu m}$ pixels).

Using $0.1\mathrm{ms}$ fixed exposure time, we capture $1200$ frames corresponding to each of the following masks: left exposed, right exposed, full aperture, and inserted perturbation (knife). Unlike the numerical study in section \ref{section: numerical study}, in the experiment we must use the same knife edge perturbation for both diff-DBFH and dOTF to ensure equivalent perturbation area for the two methods. We also capture dark and flat field images and use them to correct the raw frames. We analyze the centroid and power in each frame and filter outlier frames before summing them to produce the final intensity inputs. We use the pupil and perturbation supports as reconstruction constraints in diff-DBFH and for registration and selection of the evaluation region in dOTF, estimating them from the setup geometry (e.g. single hexagon diameter in pupil-plane pixels) and basic image analysis (e.g. sensor-to-mask rotation).

\begin{figure}[H]
    \includegraphics[width=\linewidth]{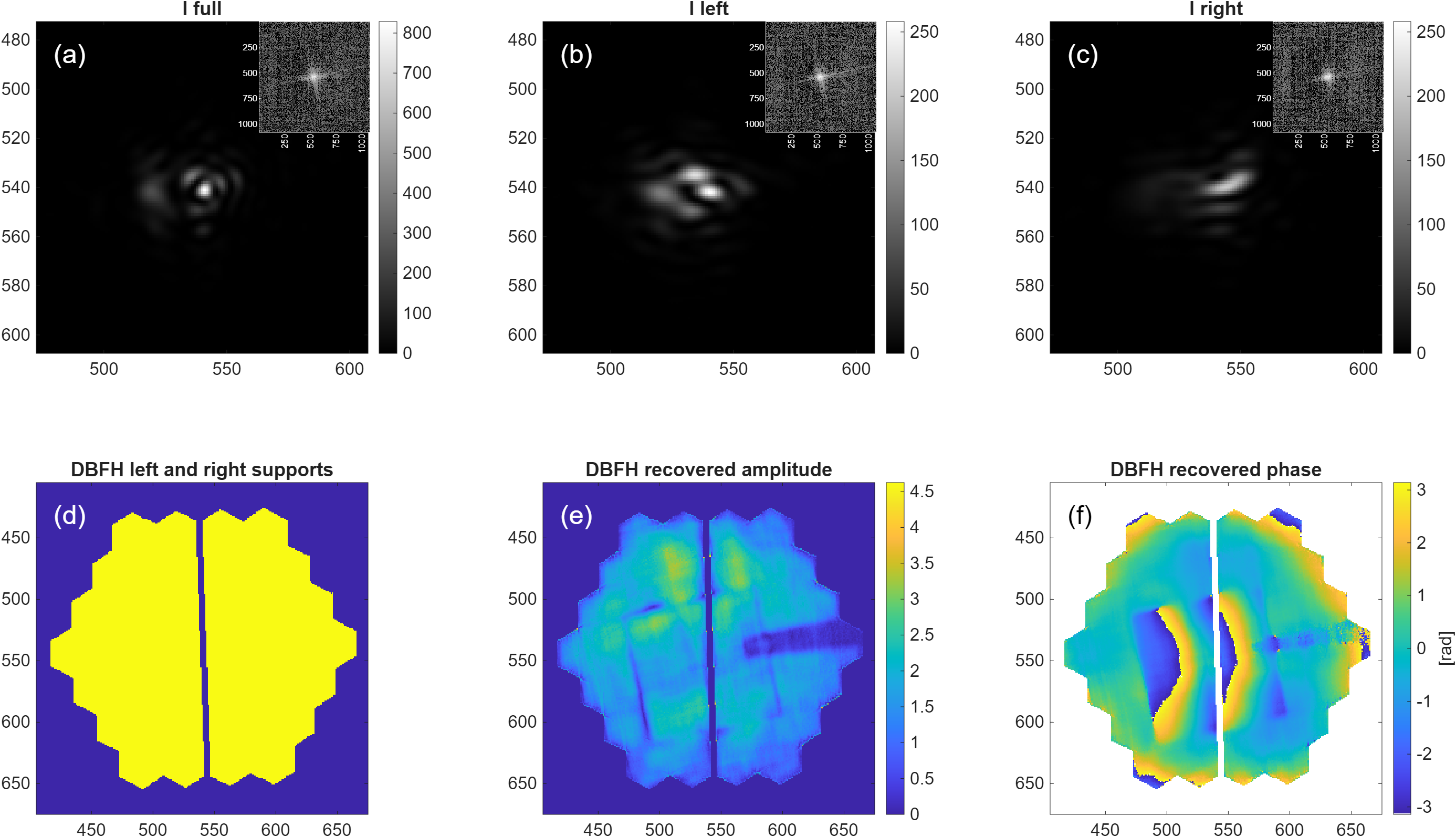}
    \vspace{1mm}
    \caption{\textbf{DBFH reference analysis.} Top row: Input intensities corresponding to full, left and right exposed sub-apertures. The insets show the full PSFs in log scale, and the larger images show a zoom-in into the central region of the PSF in linear scale. Bottom row: the left and right masks estimates input to DBFH, and the resulting pupil plane amplitude and phase reference estimates.}
    \label{fig:DBFH_reference_analysis}
\end{figure}

Figure \ref{fig:DBFH_reference_analysis} shows the intensity images inputs to DBFH, and the amplitude and phase reconstruction results that are used as reference for the ensuing comparison. 
Following the measured SNR definition in Sec.~\ref{section: numerical study},
we account for both Poisson shot noise and Gaussian background noise:
\begin{equation}
    \mathrm{SNR} =
    \frac{S_{\Omega}}
    {\sqrt{S_{\Omega} + N_\Omega\sigma_{\mathrm{bg}}^2}},
\end{equation}
where $\Omega$ is the focal plane area containing 95\% of the total image intensity, $S_{\Omega}$ is the total signal in photoelectrons within that
region after summing the frames, $N_\Omega$ is the number of
pixels in the region, and $\sigma_{\mathrm{bg}}$ is the background
noise standard deviation per pixel in the summed image.
This yields an estimated SNR of $10^4$.
Previous studies \cite{pedatzur2026double,pedatzur2026experimental} indicate that this SNR corresponds to a per-pixel phase RMSE of $\sigma_{DBFH}\leq0.1~\mathrm{rad}$, providing a sufficiently accurate reference for comparing the two perturbative methods, which are expected to be less precise.

\begin{figure}[H]
    \includegraphics[width=\linewidth]{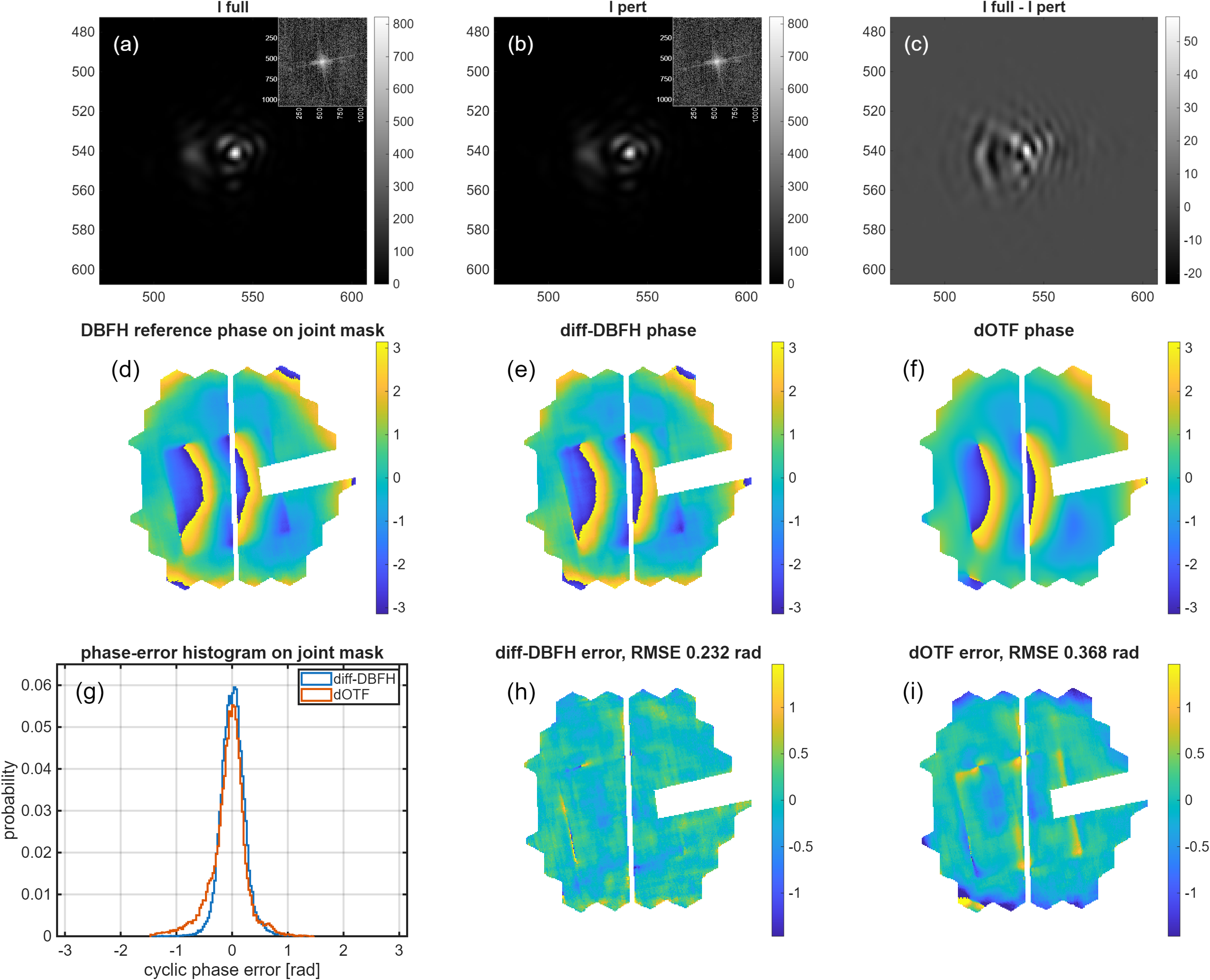}
    \caption{\textbf{Experimental comparison.} (a-b) Focal plane intensity images corresponding to the full and perturbed apertures and their difference (c) constitute the inputs to both diff-DBFH and dOTF. (d) Reference pupil plane phase from left/right DBFH.  (e-f) Phase estimates output from diff-DBFH and dOTF. Areas associated with low accuracy or low amplitude are rejected. Phase errors associated with each of the perturbative methods displayed as a histogram (g), and spatial distribution (h-i).}
    \label{fig:diff-DBFH and dOTF results}
\end{figure}

We apply the perturbation by manually actuating a vertical knife edge so as to clip the aperture by $2.85$mm, fig. \ref{fig:exp_setup and photo}(b). We obtain the perturbative and differential focal point images, fig. \ref{fig:diff-DBFH and dOTF results}(a–c), which will be used by both perturbative methods. For the purpose of this comparison we do not consider phase estimation in areas of the pupil that are invalid for either of the methods. 
Fig.~\ref{fig:diff-DBFH and dOTF results}(d–f) shows the reference (DBFH), diff-DBFH, and dOTF phase estimates over a common mask retaining 87\% of the illuminated pupil. The mask excludes regions where dOTF cannot reliably recover the pupil phase due to the finite perturbation, together with a manually defined polygon covering the low-intensity tweezers shadow, where phase is poorly defined.
Fig. \ref{fig:diff-DBFH and dOTF results}(g-i) show the cyclic phase difference relative to the reference phase in histogram form, and in the pupil plane. Using the same phase  error definition from section \ref{section: numerical study} now relative to the DBFH reference, the experimentally measured wavefront-estimation phase RMSEs are $0.232~\mathrm{rad}$ for diff-DBFH and $0.368~\mathrm{rad}$ for dOTF. In this case, the diff-DBFH least-squares solver achieved convergence after 2000 solver iterations. Solving this system repeatedly on RTX 2080ti 11Gb GPU we find a mean solve time of 32.6s with a standard deviation of 0.36s across 10 runs, each capped at 2000 iterations.

The advantage of the DBFH framework is clear: while both methods employ the same approximation (neglecting the intensity associated with the perturbation sub-aperture), dOTF's elegant yet over-simplistic readout in the autocorrelation plane relies on the convolution between the field in the perturbation sub-aperture and its exposed part. Diff-DBFH manages to reconstruct sharper features which are naturally smeared by the cross-term convolution that drives dOTF.

Note that the experimental results report better phase errors compared with the numerical study. This discrepancy is simply due to the phase pattern we chose in each case: the simulation WFE includes significantly higher spatial frequency content. It should also be noted that our numerical study did not take into account experimental conditions such as source drift, residual rotations between aperture and sensor, possible motion of the mask between captures and imperfect knowledge of the compact support (aperture).

\begin{figure}[H]
    \centering
    \includegraphics[width=0.5\linewidth]{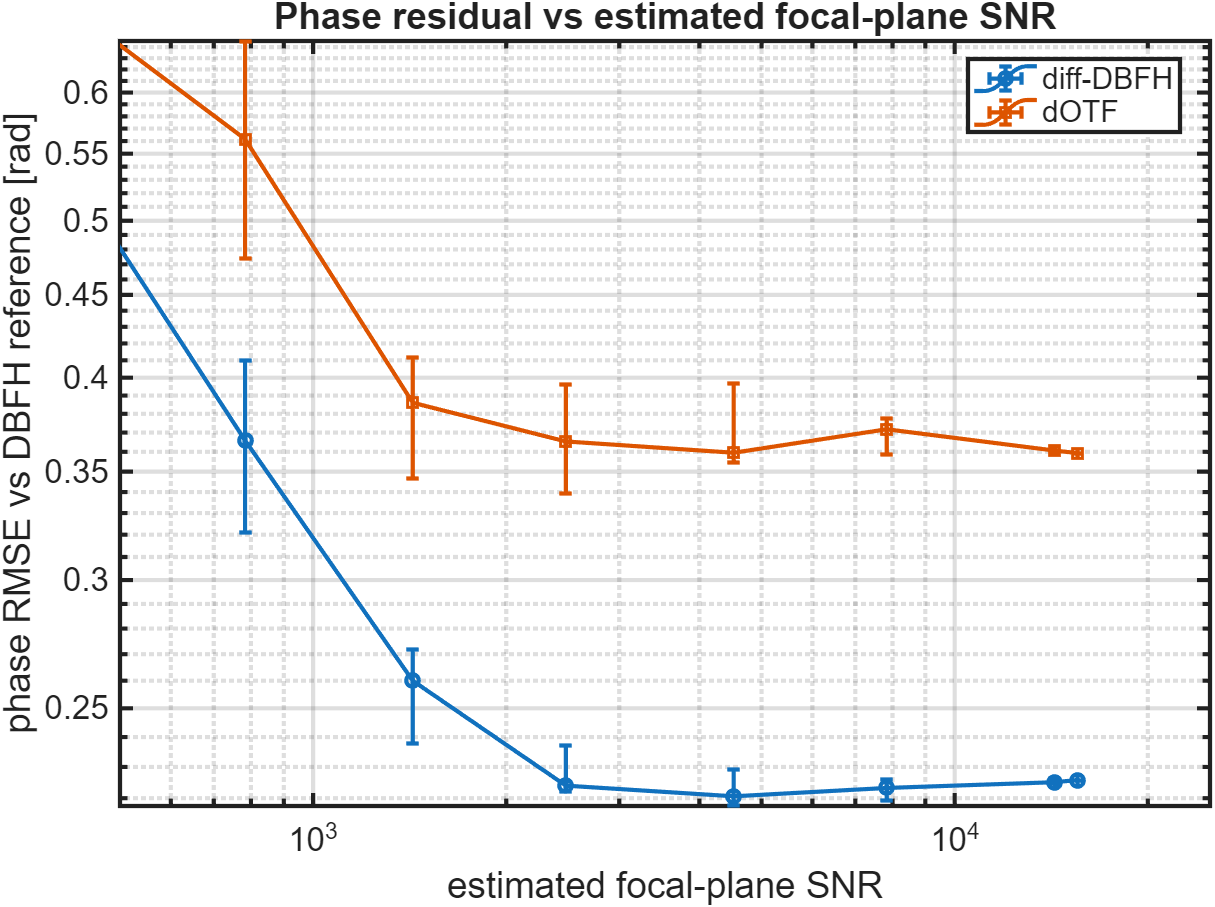}
    \vspace{1mm}
    \caption{\textbf{Experimental sweep of noise conditions.} Experimental comparison of diff-DBFH and dOTF phase RMSE relative to the reference wavefront under different noise conditions. An effective sweep of SNR is achieved by including randomly drawn subsets of frames out of the full dataset. Each SNR condition (except the full dataset) was sampled $10$ times by repeating the calculation with different combinations of randomly selected subsets of frames. Markers indicate the median measured SNR and phase RMSE across realizations. Horizontal and vertical error bars span the 25th--75th percentiles of measured SNR and phase RMSE, respectively. Horizontal error bars are smaller than the markers.}
    \label{fig:experimental Phase RMSE vs estimated SNR}
\end{figure}

Next, we wish to compare the performance of the two perturbative methods under different noise conditions experimentally, and compare it to our numerical study. We sweep the SNR by summing subsets of 1, 3, 10, 30, 100, 300, 1000, and 1200 frames, drawing multiple random subsets at each size (except 1200) to characterize statistical variability. The same trend we highlighted in fig. \ref{fig:diff-DBFH vs dOTF performance}(a) is observed in our experimental SNR sweep as well, fig. \ref{fig:experimental Phase RMSE vs estimated SNR}. Exact agreement with the simulation is not expected, primarily because the experiment involves a different perturbation size, a particular pupil-field, and includes noise sources absent from the model, such as detector readout noise.

Lastly, diff-DBFH avoids a critical limitation inherent to the standard dOTF readout. In dOTF, the pupil field is inferred from a shifted autocorrelation-plane copy rather than solved for directly. For a finite perturbing region, this copy represents a local convolution with the perturbation field, so the recovered phase is effectively smeared by the perturbation area. When the field is slowly varying inside the perturbation area, this averaging mainly behaves as a low-pass filter, as seen in fig.~\ref{fig:diff-DBFH and dOTF results}(f). However, when the perturbation overlaps sharp phase or amplitude structure, the local convolution no longer represents the underlying pupil phase reliably. While this limitation can in principle be mitigated by using multiple perturbation sites \cite{martinez2023bisymmetric}, it does not arise in diff-DBFH. In diff-DBFH, the differential intensity is used to estimate the cross term, after which the pupil field is recovered through a constrained linear inverse problem rather than by direct autocorrelation-plane readout. Thus, aside from the shared perturbative approximation associated with neglecting $I_2 = |A_2|^2$, diff-DBFH avoids the convolution smearing that limits dOTF, although it remains sensitive to measurement noise, support mismatch, and model error.

\section{Conclusions}\label{section: conclusions}

In this work, we introduced differential DBFH as a perturbative wavefront-sensing method that uses two focal-plane intensity captures to estimate the complex pupil-plane field. The method is motivated by dOTF, but avoids direct autocorrelation-plane phase readout by leveraging the algebraically linear DBFH framework. Through numerical simulations and bench-top experiments using an HWO EAC5-like aperture as a representative model system, we showed that diff-DBFH provides more accurate phase recovery than dOTF across a range of noise levels and perturbation geometries. These results suggest that the inherent limitations commonly associated with perturbative focal-plane wavefront sensing can be substantially reduced when the measurement is treated as a constrained inverse linear problem rather than a direct readout. The present formulation assumes monochromatic illumination. Assessing the effects of finite bandwidth on reconstruction accuracy remains a subject for future work.

\section{Appendix A: Estimation of $\operatorname{Im}(A_1^*A_2)$} \label{appendix: Estimation of $Im(A_1^*A_2)$}

\begin{figure}[H]
    \centering
    \includegraphics[width=1\linewidth]{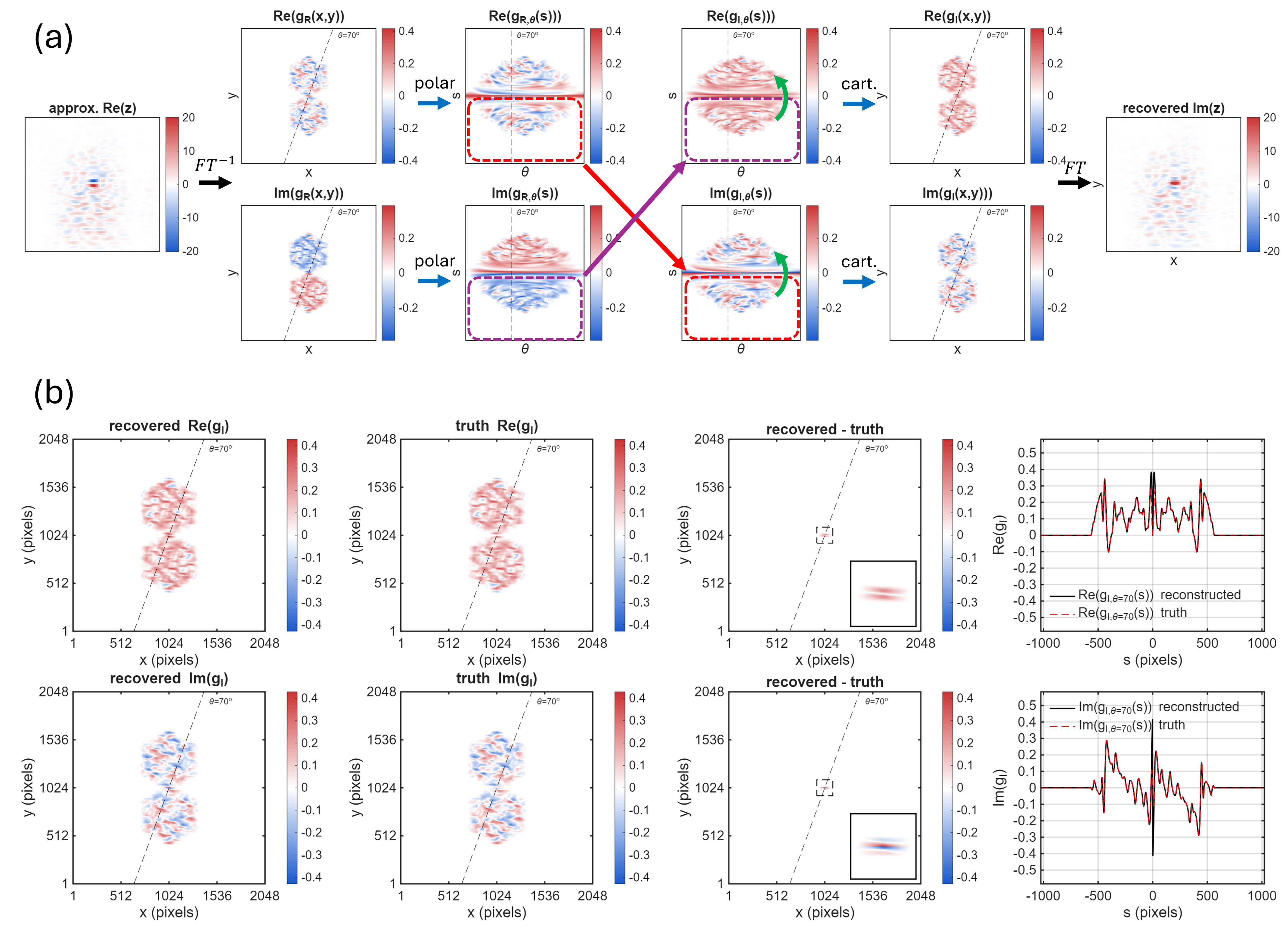}
    \caption{\textbf{Recovery of $\operatorname{Im}(A_1^{*}A_2)$ from
    $\operatorname{Re}(A_1^{*}A_2)$.} (a) Flowchart describing the principal steps of the reconstruction of $\operatorname{Im}(z)$ from $\operatorname{Re}(z)$. The perturbation geometry is chosen such that $g=g_R+i g_I = \mathcal{F}^{-1}[A_1^{*}A_2]$ vanishes over one
    half-plane. Each radial coordinate $s$ is oriented so that $s<0$ lies
    within this null half-plane, where $g_I(s)=i g_R(s)$. The values of $g_I$ for $s>0$ are then obtained using the Hermitian symmetry of $g_I$.
    The recovered polar representation is interpolated onto a Cartesian
    grid and Fourier transformed to obtain $\operatorname{Im}(z)$.
    (b) Comparison of the recovered and ground-truth real and imaginary components of $g_I$ in the autocorrelation domain, together with their residuals. Insets in the residual panels magnify the central regions marked by dashed boxes, using the same color scales as the main panels to reveal localized reconstruction errors. The rightmost panels compare the recovered and ground-truth values along a representative slice at $\theta=70^{\circ}$, providing a higher-sensitivity view of reconstruction errors that are difficult to discern in the two-dimensional residual maps.}
\label{fig:Im(z) flowchart}
\end{figure}

Figure \ref{fig:Im(z) flowchart}(a) presents an algorithmic flowchart which describes the basic conceptual steps used to recover $\operatorname{Im}(z)$ from $\operatorname{Re}(z)$, where $z=A_1^*A_2$, corresponding to the random wavefront used in the numerical study (section \ref{section: numerical study}). Starting from the approximate real part of the cross-term, $\operatorname{Re}(z)$ in the focal plane, we apply an inverse Fourier transform (black arrow) back to the pupil plane (strictly speaking, the autocorrelation domain). The resulting function, which we call $g_R$, is complex and has non-vanishing values only for combinations of vectors living in the CSs of the two fields $E_1$ and the small perturbation $E_2$. Note that $g = g_R+ig_I$ is non-vanishing only on the upper half plane (not shown in Fig. \ref{fig:Im(z) flowchart}) due to the choice of perturbation geometry. Thus, for any slice passing through the center, $g_I=ig_R$ in the lower half plane. To take advantage of this property we move to polar coordinates $(x,y)\rightarrow(s,\theta)$ (blue arrows), and `copy' the lower half planes of $g_R$ to their missing counterpart in $g_I$ (red and purple arrows). Using the Hermitian symmetry property of Fourier transforms we extend our solutions to the upper half-plane (green arrows). We interpolate our solution to $g_I$ back from polar to cartesian coordinates, followed by forward Fourier transform back to the focal plane (black arrow). 

This scheme can be elegantly implemented using the Radon transform and the slice-projection theorem. Applying Radon on $\operatorname{Re}(z)$ projects along every angle $\theta$, and a 1D Fourier transform over these projections is equivalent to a $\theta$ slice of $g_{R,\theta}$. We continue as above to reconstruct $g_{I,\theta}$ in polar coordinates and interpolate back to cartesian.

Figure \ref{fig:Im(z) flowchart}(b) compares the reconstructed imaginary part, $g_{I,\theta}$, with the simulation ground truth, focusing on an arbitrary single slice at $\theta =70^{\circ}$  passing through the origin. The errors are due to the combined effect of our approximation of $\operatorname{Re}(z)$ and the injected noise, and appear close to the origin.

\section{Appendix B: Noise and Perturbation Size Tradeoff}\label{appendix: Noise and Perturbation Size Tradeoff}
In the noiseless case where we are limited only by the validity of the approximation $\Delta I\approx2\operatorname{Re}(A_1^*A_2)$, we improve the reconstruction result by making the perturbation smaller. However, in the noisy case, the same action degrades the SNR in the differential image $\Delta I$. These tradeoffs often give rise to a `sweet-spot' in terms of perturbation sub-aperture size for every given noise level.

\begin{figure}[H]
    \includegraphics[width=1\linewidth]{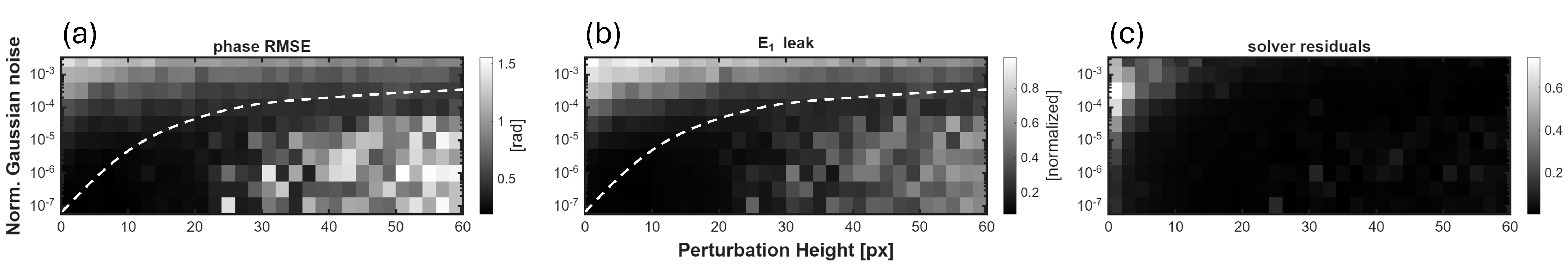}
    \vspace{1mm}
    \caption{\textbf{Noise and perturbation size tradeoff.} The residual phase error, $\phi_{RMSE}$ (a), and normalized intensity leak values outside the support of $E_1$ in the pupil plane (b), and solver residual (c) for swept combinations of perturbation vertical height and applied Gaussian noise (normalized to signal peak). The interplay between statistical noise and systematic bias gives rise to a different optimal choice of perturbation size for every noise level, as depicted by the dashed white trend line.}
    \label{fig:error diagnostics}
\end{figure}

Fig. \ref{fig:error diagnostics} depicts a sweep of these two critical factors: the horizontal axis are the growing values of perturbation size, and the vertical axis are the injected Gaussian noise levels. These figures clearly demonstrate the tradeoff discussed above between the systematic bias associated with the approximation and the effect of random measurement noise. Consider a choice of 40 px perturbation height, and $\sigma_{noise}=10^{-4}$ normalized Gaussian noise level. Simply reducing the perturbation size while fixing the noise level results in higher reconstruction errors and leaks as we move deeper into the regime dominated by noise. Conversely, improving the SNR (perhaps by longer integration) does not necessarily lead to an improved reconstruction, as the errors are now dominated by the approximation bias. In this analysis we allowed the solver a high but fixed maximal number of iterations ($10^4$) to facilitate a just comparison between the cases. The increase in phase errors for growing perturbation sizes is not due to poor solver convergence as indicated by the flat solver residuals (fig. \ref{fig:error diagnostics}(c)), and is a result of the larger systematic bias.

\section{Disclosures} 
The authors declare no conflicts of interest related to this work.

\section{Data and Code Availability Statements} 

The code used in this work, including demonstrations of the key diff-DBFH and $\operatorname{Im}(A_1^*A_2)$ estimation functionalities, is openly available at
\url{https://github.com/orenpedatzur/differential_DBFH}. The data supporting this study are available from the corresponding author upon reasonable request.

\section{Acknowledgments} 
We gratefully acknowledge funding from Avo Photonics that enabled the test setup.


\bibliography{diff_DBFH_bib}   

@article{codona2013differential,
  title={Differential optical transfer function wavefront sensing},
  author={Codona, Johanan L},
  journal={Optical Engineering},
  volume={52},
  number={9},
  pages={097105--097105},
  year={2013},
  publisher={Society of Photo-Optical Instrumentation Engineers}
}

@inproceedings{codona2012theory,
  title={Theory and application of differential OTF (dOTF) wavefront sensing},
  author={Codona, Johanan L},
  booktitle={Adaptive Optics Systems III},
  volume={8447},
  pages={2188--2199},
  year={2012},
  organization={SPIE}
}

@article{martinez2023bisymmetric,
  title={Bisymmetric pupil modification deconvolution strategy for differential optical transfer function (dOTF) wavefront sensing},
  author={Martinez, P and Dharmadhikari, Radhika},
  journal={Astronomy \& Astrophysics},
  volume={680},
  pages={L12},
  year={2023},
  publisher={EDP Sciences}
}

@article{martinez2024differential,
  title={Differential optical transfer function (dOTF) broad-spectrum wavefront sensing using integral field unit spectroscopy},
  author={Martinez, P},
  journal={Astronomy \& Astrophysics},
  volume={684},
  pages={L5},
  year={2024},
  publisher={EDP Sciences}
}

@article{martinez2025revisiting,
  title={Revisiting the differential optical transfer function wavefront sensing technique for high-contrast imaging},
  author={Martinez, P and Spang, A and Sallard, C and Beaulieu, M},
  journal={Astronomy \& Astrophysics},
  volume={700},
  pages={A157},
  year={2025},
  publisher={EDP Sciences}
}

@inproceedings{chanan1986segment,
  title={Segment alignment for the Keck telescope primary mirror},
  author={Chanan, Gary A and Nelson, Jerry E and Mast, Terry S},
  booktitle={Advanced Technology Optical Telescopes III},
  volume={628},
  pages={466--471},
  year={1986},
  organization={SPIE}
}

@article{gonsalves1982phase,
  title={Phase retrieval and diversity in adaptive optics},
  author={Gonsalves, Robert A},
  journal={Optical Engineering},
  volume={21},
  number={5},
  pages={829--832},
  year={1982},
  publisher={SPIE}
}

@inproceedings{dean2006phase,
  title={Phase retrieval algorithm for JWST flight and testbed telescope},
  author={Dean, Bruce H and Aronstein, David L and Smith, J Scott and Shiri, Ron and Acton, D Scott},
  booktitle={Space telescopes and instrumentation I: optical, infrared, and millimeter},
  volume={6265},
  pages={314--330},
  year={2006},
  organization={SPIE}
}

@inproceedings{gonsalves2018phase,
  title={Phase diversity: math, methods and prospects, including sequential diversity imaging},
  author={Gonsalves, Robert A},
  booktitle={Unconventional Optical Imaging},
  volume={10677},
  pages={335--345},
  year={2018},
  organization={SPIE}
}

@article{dean2003diversity,
  title={Diversity selection for phase-diverse phase retrieval},
  author={Dean, Bruce H and Bowers, Charles W},
  journal={JOSA A},
  volume={20},
  number={8},
  pages={1490--1504},
  year={2003},
  publisher={Optica Publishing Group}
}

@article{haffert2022phasing,
  title={Phasing the Giant Magellan Telescope with the holographic dispersed fringe sensor},
  author={Haffert, Sebastiaan Y and Close, Laird M and Hedglen, Alexander D and Males, Jared R and Kautz, Maggie and Bouchez, Antonin H and Demers, Richard and Quir{\'o}s-Pacheco, Fernando and Sitarski, Breann N and Van Gorkom, Kyle and others},
  journal={Journal of Astronomical Telescopes, Instruments, and Systems},
  volume={8},
  number={2},
  pages={021513--021513},
  year={2022},
  publisher={Society of Photo-Optical Instrumentation Engineers}
}

@article{cheetham2012fizeau,
  title={Fizeau interferometric cophasing of segmented mirrors},
  author={Cheetham, Anthony C and Tuthill, Peter G and Sivaramakrishnan, Anand and Lloyd, James P},
  journal={Optics express},
  volume={20},
  number={28},
  pages={29457--29471},
  year={2012},
  publisher={Optica Publishing Group}
}

@inproceedings{feinberg2007trl,
  title={TRL-6 for JWST wavefront sensing and control},
  author={Feinberg, Lee D and Dean, Bruce H and Aronstein, David L and Bowers, Charles W and Hayden, William and Lyon, Richard G and Shiri, Ron and Smith, J Scott and Acton, D Scott and Carey, Larkin and others},
  booktitle={UV/Optical/IR Space Telescopes: Innovative Technologies and Concepts III},
  volume={6687},
  pages={67--90},
  year={2007},
  organization={SPIE}
}

@article{pedatzur2019double,
  title={Double-blind holography of attosecond pulses},
  author={Pedatzur, O and Trabattoni, A and Leshem, B and Shalmoni, H and Castrovilli, MC and Galli, M and Lucchini, M and M{\aa}nsson, E and Frassetto, F and Poletto, L and others},
  journal={Nature Photonics},
  volume={13},
  number={2},
  pages={91--95},
  year={2019},
  publisher={Nature Publishing Group UK London}
}

@article{leshem2016direct,
  title={Direct single-shot phase retrieval from the diffraction pattern of separated objects},
  author={Leshem, Ben and Xu, Rui and Dallal, Yehonatan and Miao, Jianwei and Nadler, Boaz and Oron, Dan and Dudovich, Nirit and Raz, Oren},
  journal={Nature communications},
  volume={7},
  number={1},
  pages={10820},
  year={2016},
  publisher={Nature Publishing Group UK London}
}

@article{raz2014direct,
  title={Direct phase retrieval in double blind Fourier holography},
  author={Raz, Oren and Leshem, Ben and Miao, Jianwei and Nadler, Boaz and Oron, Dan and Dudovich, Nirit},
  journal={Optics express},
  volume={22},
  number={21},
  pages={24935--24950},
  year={2014},
  publisher={Optica Publishing Group}
}

@article{raz2011vectorial,
  title={Vectorial phase retrieval for linear characterization of attosecond pulses},
  author={Raz, Oren and Schwartz, Osip and Austin, Dane and Wyatt, AS and Schiavi, Andrea and Smirnova, Olga and Nadler, Boaz and Walmsley, Ian A and Oron, Dan and Dudovich, Nirit},
  journal={Physical review letters},
  volume={107},
  number={13},
  pages={133902},
  year={2011},
  publisher={APS}
}

@article{raz2013vectorial,
  title={Vectorial phase retrieval of 1-D signals},
  author={Raz, Oren and Dudovich, Nirit and Nadler, Boaz},
  journal={IEEE Transactions on Signal Processing},
  volume={61},
  number={7},
  pages={1632--1643},
  year={2013},
  publisher={IEEE}
}

@article{pedatzur2026double,
  title={Double blind Fourier holography for high-precision wavefront sensing in segmented mirror telescopes},
  author={Pedatzur, Oren},
  journal={Optics Express},
  volume={34},
  number={5},
  pages={8223--8234},
  year={2026},
  publisher={Optica Publishing Group}
}

@article{pedatzur2026experimental,
  title={Experimental Validation of Double-Blind Fourier Holography in an HWO-Like Segmented Aperture},
  author={Pedatzur, Oren and Schwartz, Noah and Acton, D Scott and Buky, Ben},
  journal={arXiv preprint arXiv:2610.34760},
  year={2026}
}

@article{feinberg2026habitable,
  title={Habitable Worlds Observatory's Concept and Technology Maturation: Initial Feasibility and Trade Space Exploration},
  author={Feinberg, Lee D and Sitarski, Breann N and McElwain, Michael W and Arney, Giada and Baker, Caleb and Bolcar, Matthew R and Levine, Marie and Liu, Alice and Mennesson, Bertrand and Roberge, Aki and others},
  journal={arXiv preprint arXiv:2601.11803},
  year={2026}
}

@book{lay2015linear,
  title={Linear Algebra and Its Applications},
  author={Lay, David C. and Lay, Steven R. and McDonald, Judi J.},
  edition={5th},
  year={2015},
  publisher={Pearson},
  address={Boston},
  note={Section 6.5, Theorem 14}
}
\bibliographystyle{spiejour}   


\vspace{2ex}\noindent\textbf{Oren Pedatzur} is an adjunct professor in the Department of Physics at Villanova University. He received a B.Sc. in physics and mathematics from the Technion, Israel, in 2009 and a Ph.D. in physics from the Weizmann Institute of Science in 2016. His doctoral research on attosecond dynamics in atoms and molecules earned him the Daniel Brenner Memorial Prize. He subsequently worked as an electro-optics engineer at Apple, where he developed the rear-facing LiDAR modules now deployed in more than 100 million iPhone, iPad and Vision Pro devices worldwide. He also served as CTO of an early-stage medical-optics startup developing a noncontact ophthalmic diagnostic system. His current research focuses on wavefront sensing for segmented-mirror space telescopes, with particular emphasis on the Habitable Worlds Observatory.

\vspace{2ex}\noindent\textbf{David T. Chuss} is Professor of Physics and Chair of the Department of Physics at Villanova University. He earned a B.S. in Physics from Villanova University in 1995, an M.S. in Physics from The Pennsylvania State University in 1997, and a Ph.D. in Physics from Northwestern University in 2002. Prior to joining the Villanova faculty in 2015, he served as an Astrophysicist at NASA's Goddard Space Flight Center following a National Research Council Research Associateship.

Dr. Chuss's research focuses on experimental astrophysics, with an emphasis on astronomical polarimetry and the development of instrumentation for observations spanning far-infrared through millimeter wavelengths. His work integrates polarization-sensitive detectors, innovative measurement techniques, and novel optical components to address fundamental questions in cosmology and the interstellar medium. He has authored more than 80 refereed journal articles, including publications in The Astrophysical Journal and Applied Optics, and is co-author of the book Astrophysical Polarimetry: Far-infrared through Millimeter Wavelengths.

He has held leadership roles in numerous national and international collaborations, including the Cosmology Large Angular Scale Surveyor (CLASS), HAWC+ on the Stratospheric Observatory for Infrared Astronomy (SOFIA), and other projects developing next-generation polarimetric instrumentation. His research has been supported by NASA and the National Science Foundation and has contributed to advances in detector technology, polarization modulation, calibration techniques, and astronomical data analysis. 

\end{spacing}
\end{document}